# Observation of high-order polarization-locked vector solitons in a fiber laser

D. Y. Tang, H. Zhang, L. M. Zhao, and X. Wu

School of Electrical and Electronic Engineering, Nanyang Technological University, Singapore 639798

We report on the experimental observation of a novel type of polarization locked vector soliton in a passively mode-locked fiber laser. The vector soliton is characterized by that not only the two orthogonally polarized soliton components are phase locked, but also one of the components has a double-humped intensity profile. Multiple such phase-locked high order vector solitons with identical soliton parameters and harmonic mode-locking of the vector solitons were also obtained in the laser. Numerical simulations confirmed the existence of stable high-order vector solitons in fiber lasers.

Soliton as a stable localized nonlinear wave has been observed in various physical systems and been extensively studied [1, 2]. Optical solitons were first experimentally observed in single mode fibers (SMFs) by Mollenauer et al. in 1980 [3]. It was shown that dynamics of the solitons could be well described by the nonlinear Schrödinger equation, a paradigm equation governing optical pulse propagation in ideal SMFs. However, in reality a SMF always supports two orthogonal polarization modes. Taking fiber birefringence into account, it was later found that depending on the strength of fiber birefringence, different types of vector solitons, such as the group velocity locked vector solitons [4-5], the rotating polarization vector solitons [6-7], and the phase locked vector solitons [8-10], could also be formed in SMFs.

Optical solitons were also observed in mode-locked fiber lasers. Pulse propagation in a fiber laser cavity is different from that in a SMF. Apart from propagating in the fibers that form the laser cavity, a pulse propagating in a laser also subjects to actions of the laser gain and other cavity components. Dynamics of solitons formed in a fiber laser is governed by the Ginzburg-Landau equation, which takes account of not only the fiber dispersion and Kerr nonlinearity, but also the laser gain and losses. However, it was shown that under suitable conditions solitons formed in fiber lasers have analogous features to those of solitons formed in SMFs. Furthermore, vector solitons were also predicted in mode-locked fiber lasers and confirmed experimentally recently [11-12].

Among the various vector solitons formed in mode-locked fiber lasers or SMFs, the phase locked one has attracted considerable attention. Back to 1988 Christodoulides and Joseph first theoretically predicted a novel form of phase locked vector soliton in birefringent dispersive media [8], which is now known as a high order phase locked vector soliton in SMFs. The fundamental form of the phase locked temporal vector solitons was recently experimentally observed [12-13]. However, to the best of our knowledge, no high order temporal vector solitons have been demonstrated. Numerical studies have shown that the high order phase locked vector solitons are unstable in SMFs [7]. In this letter, we report on the experimental observation of a stable phase-locked high order vector soliton in a mode-locked fiber laser. Multiple high order vector solitons with identical soliton parameters coexisting in laser cavity and harmonic mode-locking of the high order vector solitons were also observed. Moreover, based on a coupled Ginzburg-Landau equation model we show numerically that phase locked high order vector solitons are stable in mode-locked fiber lasers.

The experimental setup is shown in Fig. 1. The fiber laser has a ring cavity consisting of a piece of 4.6 m Erbium-doped fiber (EDF) with group velocity dispersion parameter 10 ps/km/nm and a total length of 5.4 m standard single mode fiber (SMF) with group velocity dispersion parameter 18 ps/km/nm. Mode-locking of the laser is achieved with a semiconductor saturable absorption mirror (SESAM). Note that within one cavity round-trip the pulse propagates twice in the SMF between the circulator and the SESAM. A polarization independent circulator was used to force the unidirectional operation of the ring and simultaneously to incorporate the SESAM in the cavity. The laser was pumped by a high power Fiber Raman Laser source (BWC-FL-1480-1) of wavelength 1480 nm. A 10% fiber coupler was used to output the signals. The SESAM used is made based on GaInNAs quantum wells. It has a saturable absorption modulation depth of 5%, a saturation fluence of 90 μJ/cm$^2$ and a recovery time of 10 ps. The central absorption wavelength of the SESAM is at 1550nm.

As no polarizer was used in the cavity, depending on the net cavity linear birefringence, various types of vector solitons such as the group velocity locked vector solitons, the polarization rotating vector solitons, and the fundamental phase locked vector solitons were obtained in the laser. Especially, we found that the experimentally observed features of these vector solitons could be well described by an extended coupled Ginzburg-Landau equation model, which also considered effects of the saturable absorber and the laser cavity [14]. Encouraged by the results we had further searched for the high order phase locked vector solitons theoretically predicted. Through splicing a fiber pigtailed optical isolator between the output port and the external cavity measurement apparatus, which serves as suppressing the influence of spurious back reflection on the laser operation, we could indeed obtain one of such vector solitons. Fig.2 shows for example the optical spectra and autocorrelation traces of the soliton. Polarization locking of the soliton is identified by measuring the polarization evolution frequency (PEF) of the soliton pulse train [13]. No PEF could be detected. As the vector soliton has a stationary elliptic polarization, we could use an external polarizer to separate its two orthogonal polarization components. The optical spectra of the components are shown in Fig. 2a. The spectra have the same central wavelength and about 10dB peak spectral intensity difference. Both spectra display soliton sidebands. It shows that both of the components are optical soliton. In addition, coherent energy exchange between the two soliton components, represented by the appearance of spectral peak-dip sidebands [14], is also visible on the spectra. Different from the polarization resolved spectra of the fundamental phase locked vector solitons, there is a strong spectral dip at the center of the soliton spectrum of the weak component, while no such dip on the spectrum of the strong soliton component. To identify the formation mechanism of the spectral dip, we further measured the autocorrelation traces of each of the soliton components. It turned out that the weak component of the vector soliton had a double-humped intensity profile as shown in Fig. 2b. The pulse width of the humps is about 719 fs if the Sech$^2$ profile is assumed, and the separation between the humps is about 1.5 ps. The strong component of the vector soliton is a single-hump soliton. It has a pulse width of about 1088 fs if the Sech$^2$ profile is assumed. The components of the vector soliton have the pulse intensity profiles exactly as those predicted by Akhmediev et al [ 10 ] and Christodoulides [8] for a high order phase locked vector soliton. Furthermore, the spectral dip at the center of the spectrum indicates that the two humps have 90° phase difference, which is also in agreement with the theoretical prediction.

Once the laser operation conditions were appropriately selected, the high order phase locked vector soliton operation was always obtained in the laser. Experimentally, multiple such vector solitons with identical soliton parameters were also obtained. Through carefully changing the pump strength one could even control the number of the vector solitons in cavity, and it didn't change the structure of the vector solitons. Like the scalar solitons observed in the conventional soliton fiber lasers, harmonic mode locking of the high order phase locked vector solitons was also observed, as shown in Fig. 3, where 8 such vector solitons were equally spaced in the cavity. All our experimental results show that formation of the high order phase locked vector solitons is an intrinsic feature of the fiber laser.

To confirm our experimental observations, we also numerically simulated the operation of the laser. We used the following coupled Ginzburg-Landau equations to describe the pulse propagation in the weakly birefringent fibers in the cavity:

$$\begin{cases} \frac{\partial u}{\partial z} = i\beta u - \delta\frac{\partial u}{\partial t} - \frac{ik''}{2}\frac{\partial^2 u}{\partial t^2} + \frac{ik'''}{6}\frac{\partial^3 u}{\partial t^3} + i\gamma(|u|^2 + \frac{2}{3}|v|^2)u + \frac{i\gamma}{3}v^2 u^* + \frac{g}{2}u + \frac{g}{2\Omega_g^2}\frac{\partial^2 u}{\partial t^2} \\ \frac{\partial v}{\partial z} = -i\beta v + \delta\frac{\partial v}{\partial t} - \frac{ik''}{2}\frac{\partial^2 v}{\partial t^2} + \frac{ik'''}{6}\frac{\partial^3 v}{\partial t^3} + i\gamma(|v|^2 + \frac{2}{3}|u|^2)v + \frac{i\gamma}{3}u^2 v^* + \frac{g}{2}v + \frac{g}{2\Omega_g^2}\frac{\partial^2 v}{\partial t^2} \end{cases} \quad (1)$$

Where, u and v are the normalized envelopes of the optical pulses along the two orthogonal polarized modes of the optical fiber. $2\beta = 2\pi\Delta n/\lambda$ is the wave-number difference between the two modes. $2\delta = 2\beta\lambda/2\pi c$ is the inverse group velocity difference. $k''$ is the second order dispersion coefficient, $k'''$ is the third order dispersion coefficient and $\gamma$ represents the nonlinearity of the fiber. g is the saturable gain coefficient of the fiber and $\Omega_g$ is the bandwidth of the laser gain. For undoped fibers g=0; for erbium doped fiber, we considered its gain saturation as

$$g = G\exp[-\frac{\int(|u|^2 + |v|^2)dt}{P_{sat}}] \quad (2)$$

where G is the small signal gain coefficient and $P_{sat}$ is the normalized saturation energy.

The saturable absorption of the SESAM is described by the rate equation:

$$\frac{\partial l_s}{\partial t} = -\frac{l_s - l_0}{T_{rec}} - \frac{|u|^2 + |v|^2}{E_{sat}}l_s \quad (3)$$

Where $T_{rec}$ is the absorption recovery time, $l_0$ is the initial absorption of the absorber, and $E_{sat}$ is the absorber saturation energy. We used the following parameters for our simulations for possibly matching the experimental conditions: $\gamma=3$ $W^{-1}km^{-1}$, $\Omega_g =24$ nm,

$P_{sat}$=50 pJ, $k''_{SMF}$=-23 ps$^2$/km, $k''_{EDF}$=-13 ps$^2$/km, $k'''$=-0.13 ps$^3$/km, $E_{sat}$=10 pJ, $l_0$=0.3, and $T_{rec}$ = 2 ps, cavity length $L$= 10 m.

We used the standard split-step Fourier technique to solve the equations and a so-called pulse tracing method to model the effects of laser oscillation [15]. We have always started our simulations with an arbitrary weak light input. Fig. 4 shows one of the typical results obtained. With a cavity linear birefringence of $L_b$=3$L$, a stable high order phase locked vector soliton state was obtained. The weak polarization component of the vector soliton consists of two bound solitons with pulse separation of ~ 1 ps, while the strong polarization component of the vector soliton is a single-hump soliton. It is interesting to see that the pulse of the strong component is only temporally overlapped with one of the two pulses of the weak component. Due to the strong cross-phase coupling between the temporally overlapped pulses, the two pulses of the weak components have different pulse widths and intensities. Propagating along the cavity, obvious coherent energy exchange between the two temporally overlapped solitons is visible. Fig 4b further gives the calculated spectra of the vector soliton components, which also show that the phase difference between the two bound solitons of the weak componet is 90°.

Depending on the laser parameter selections, other high order phase locked vector soltions, such as the one with both soliton components having a double-hunmped structure, were also numerically obtained. We note that similar high order vector solitons were also predicted for pulse propagation in weakly birefringent fibers, but they are unstable. However, we found that all the numerically obtained high order phase locked vector solitons were stable in the laser. We believe that the different stability feature of the high order phase loceked vector solitons in fiber and in fiber lasers could be traced back to their different soliton nature. While the soliton formed in a SMF is essentially a hamiltonian soltion, the one formed in a fiber laser is a dissipative soliton, which is in fact a strang attactor of the laser system. The formation of multiple identical high-order vectore solitons in the fiber laser clearly show the dissipative nature of the formed vector solitons.

In conclusion, we have first experimentally observed a novel type of high order phase locked vector soliton in a passively mode-locked fiber laser. The high order vector soliton is characterized by that its two orthogonal polarization components are phase locked, and while the stronger polarization component is a single hump pulse, the weaker component has a double-humped structure with 90° phase difference between the humps. Our experimental result firstly confirmed the theoretical predictions on the high order phase locked vector solitons in birefringent dispersive media.

**Figure captions:**

Fig.1: Schematic of the vector soliton fiber laser. WDM: wavelength division multiplexer. EDF: erbium doped fiber.

Fig. 2: Polarization resolved soliton spectra and autocorrelation traces of the vector soliton observed. (a) Soliton spectra. (b) Autocorrelation traces.

Fig. 3: Oscilloscope trace of a harmonically mode-locked high order phase locked vector soliton state. Lc: cavity roundtrip time. 8 vector solitons coexist in cavity.

Fig. 4: A stable high order phase locked vector soliton state numerically calculated. (a) Soliton intensity profiles of the two orthogonally polarized components. (b) The corresponding optical spectra of (a).

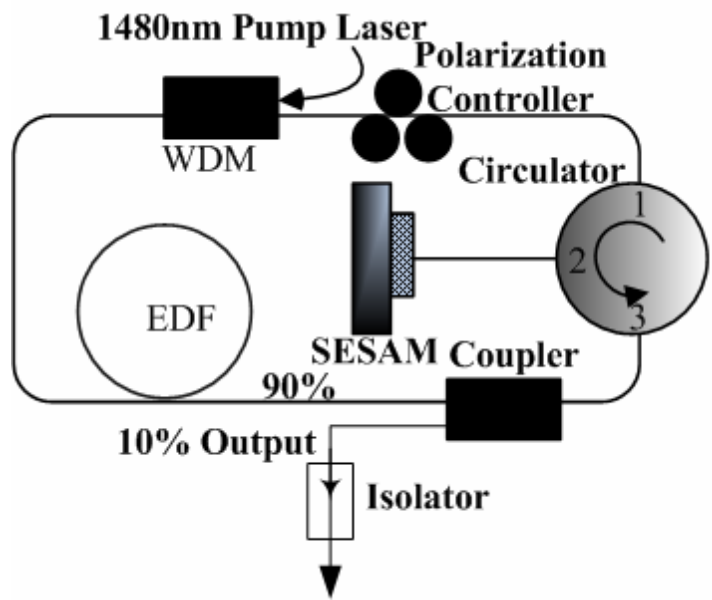

Fig. 1. D. Y. Tang et al.

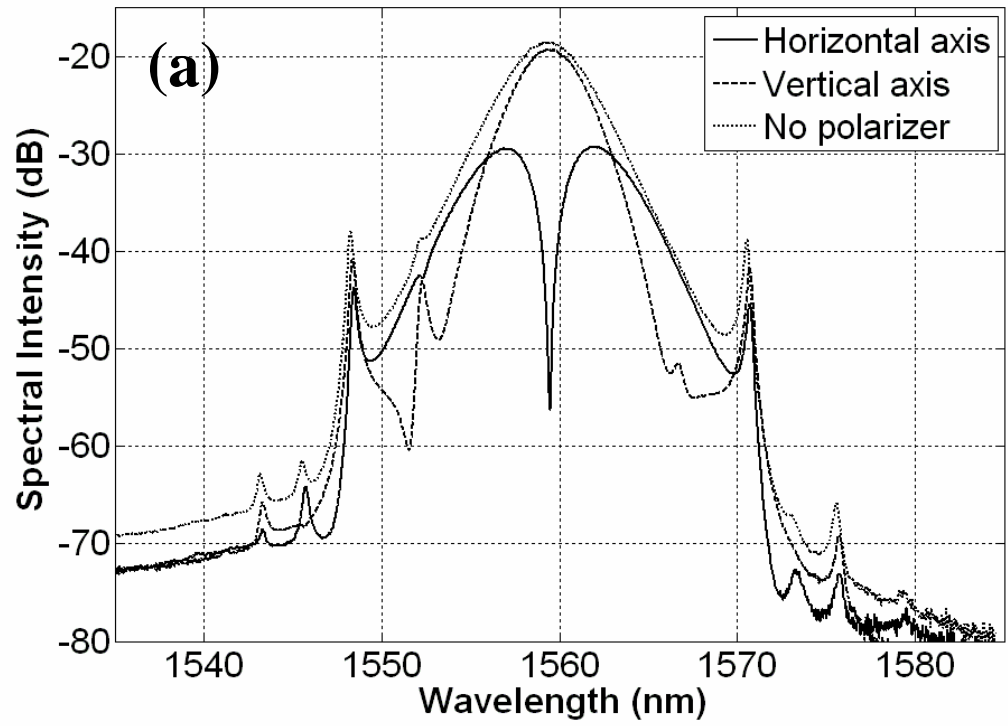

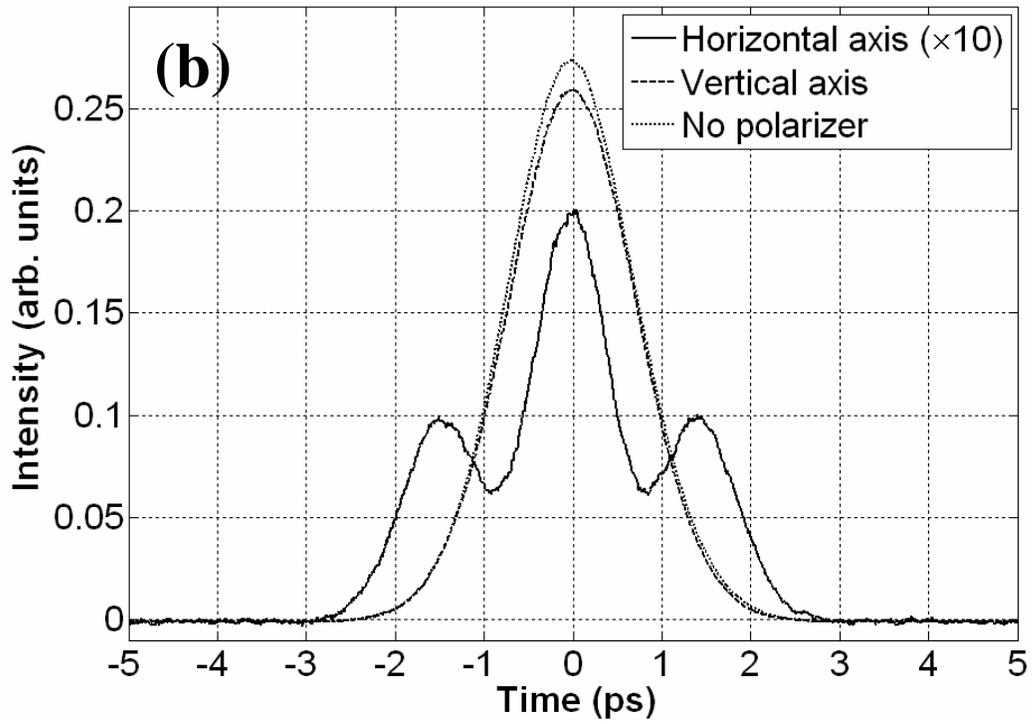

Fig. 2. D. Y. Tang et al.

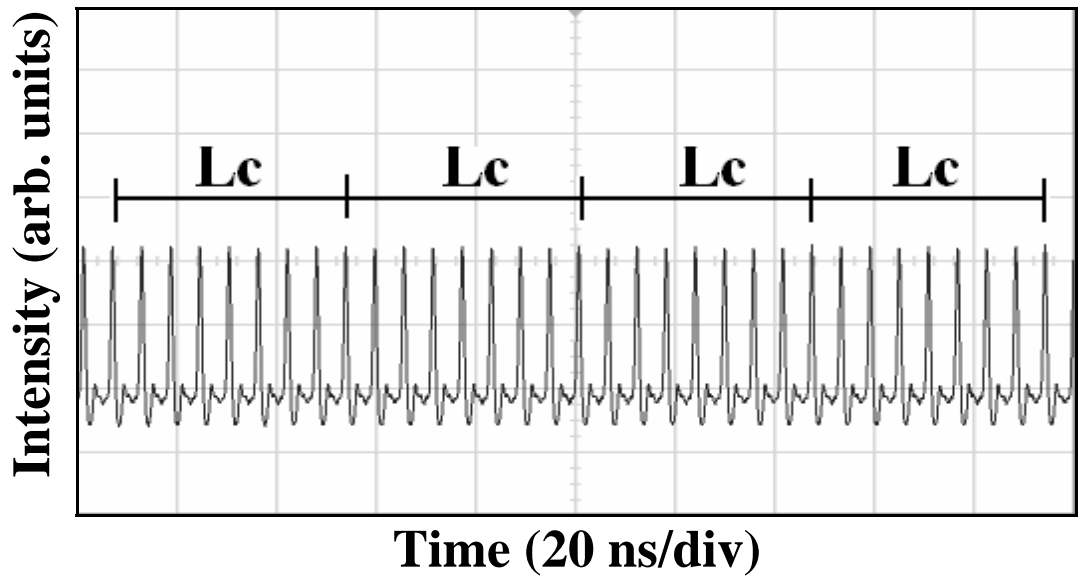

Fig. 3. D. Y. Tang et al.

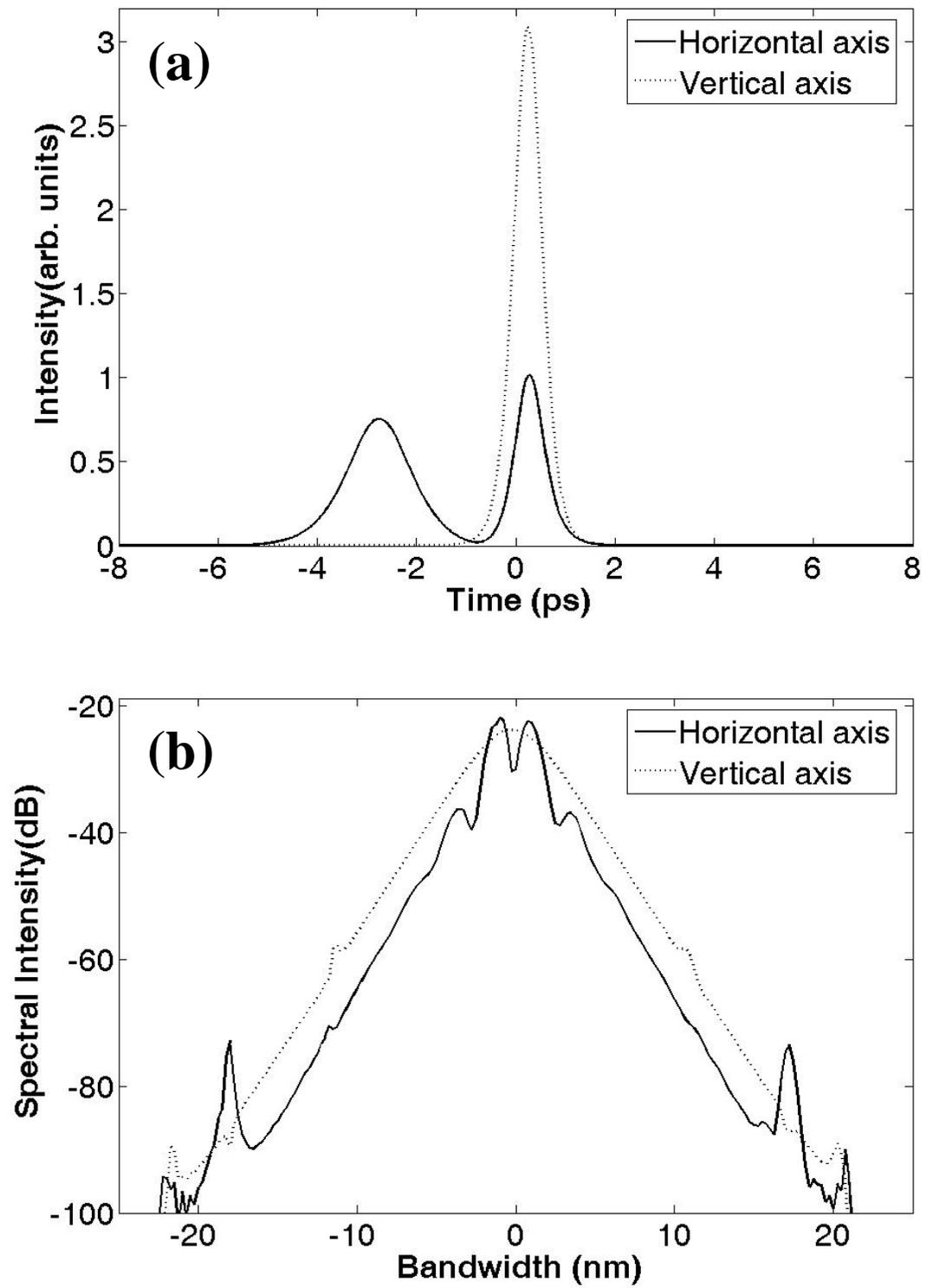

Fig. 4. D. Y. Tang et al.